# Long Distance Signaling Using Axion-like Particles

Daniel D. Stancil, Department of Electrical and Computer Engineering

Carnegie Mellon University, Pittsburgh, PA 15213

*Abstract*

The possible existence of axion-like particles could lead to a new type of long distance communication. In this work, basic antenna concepts are defined and a Friis-like equation is derived to facilitate long-distance link calculations. An example calculation is presented showing that communication over distances of 1000 km or more may be possible for $m_a < 3.5$ meV and $g_{a\gamma\gamma} > 5\times10^{-8}$ GeV$^{-1}$.

The axion has been proposed as a solution to the strong-CP problem [1], and is also a candidate for the galactic dark matter [2]. Interest in axions has increased recently owing to the report by the PVLAS collaboration of optical rotation induced by a magnetic field in a vacuum [3], since the creation of axions or other similar particles was one possible explanation for this rotation. Subsequently a number of groups began experimental searches using axion generation and detection schemes that would more definitively point to these new particles as the explanation [4]. The PVLAS result was surprising because it suggested coupling between the axion and electromagnetic fields that was much larger than thought possible based on solar axion observations by the CAST collaboration [5]. Although mechanisms have been proposed to reconcile the reports [6], the PVLAS collaboration recently retracted the results [7] and an independent group has reported a negative result from a photon regeneration experiment that excludes the PVLAS result [8].

There does not now appear to be any experimental evidence of a coupling strength inconsistent with CAST observations. However, since recent work has suggested mechanisms whereby such strong coupling may be possible, I believe it remains interesting to consider the implications of stronger-than-expected axion-photon coupling. In particular, I would like to call attention to the observation that a new type of long-distance signaling and communication may be possible. It may be possible to construct a communication system that cannot be blocked—even communicating directly through the diameter of the earth. This would make reliable worldwide signaling possible without the use of either satellites or the ionosphere, and would enable communication to locations previously inaccessible, such as submarines at the bottom of the sea and mines deep beneath the earth. The signal would also be very difficult to intercept since the axion beam would be essentially as narrow as a laser beam used to create it, and most of the path would be underground. With advances in power and sensitivity, it may also be possible to use axion signaling in space communications. For example, using such a system, communication with points on the far side of the moon may be possible without the use of lunar satellites.

Communication systems using neutrinos have also been proposed [9], and would have many of the same characteristics as the proposed axion system. However, the generation and detection of neutrinos requires massive particle accelerators and scintillation detectors [10]. Also, full deflection over 4π steradians would not be practical, though limited beam steering could be achieved using a magnetic field to deflect the precursor pion beam. Finally, it would be difficult to consider modulation techniques more sophisticated than simple amplitude modulation. In contrast, using axion mass and coupling values not yet experimentally explored, it appears that worldwide communication would be possible with a fully steerable axion system about the size of a medium-size telescope. Further, since the signals at the input and output would be electromagnetic waves, any existing modulation technology could be used.

However, as with neutrinos, the lack of strong interactions with matter presents challenges with respect to the generation and detection of axions. Sikivie proposed an experimental approach for detecting axions via their coupling to the electromagnetic field [11]. The coupling was obtained by considering the Lagrangian density

$$L = -\frac{1}{4}F_{\mu\nu}F^{\mu\nu} - \frac{1}{4}g_{a\gamma\gamma}aF_{\mu\nu}\tilde{F}^{\mu\nu} + \frac{1}{2}\partial_\mu a \partial^\mu a - \frac{1}{2}m_a^2 a^2, \qquad (1)$$

where $F_{\mu\nu} = \partial_\mu A_\nu - \partial_\nu A_\mu$ is the electromagnetic field tensor, $A_\alpha = (V, -\vec{A})$ where $V$ is the electric scalar potential and $\vec{A}$ is the magnetic vector potential, $a$ is the axion field, $g_{a\gamma\gamma}$ is the coupling constant between the electromagnetic and axion fields, and the electromagnetic dual tensor is given by $\tilde{F}_{\alpha\beta} = \frac{1}{2}\varepsilon_{\alpha\beta\gamma\delta}F^{\gamma\delta}$. In these equations we have taken $\hbar = c = 1$.

As an example, consider the coupling between plane waves propagating along the $z$ direction caused by a strong static magnetic field parallel to the polarization of the incident electromagnetic wave. If the time dependence is $\exp(-i\omega t)$, where $\omega$ is the frequency of the incident linearly polarized electromagnetic wave, the equations of motion obtained from (1) reduce to the coupled equations

$$\frac{\partial^2 a}{\partial z^2} + \left(\omega^2 - m_a^2\right)a = -i\omega g_{a\gamma\gamma} B_0 A, \qquad \frac{\partial^2 A}{\partial z^2} + \omega^2 A = i\omega g_{a\gamma\gamma} B_0 a. \tag{2}$$

Thus the static magnetic field $B_0$ couples the photon and axion fields. An apparatus for the generation and detection of axions based on this coupling is shown in Figure 1(a). This is sometimes referred to as an "invisible light through walls" experiment [4,12,13]. An electromagnetic wave with amplitude $A(0)$ enters a region of magnetic field of strength $B_{OT}$ extending over a distance $L_T$. If the coupling is sufficiently weak that the change in $A$ over the transmit conversion region and the change in $a$ over the receive conversion region are negligible, then the conversion loss through the system will be [12]

$$P_0 / P_{in} = p_R p_T, \tag{3}$$

where $P_{in}$ is the optical input power, $P_0$ is the optical output power, $p_T, p_R$ are the probabilities of photon-axion (and axion-photon) conversion in the transmitter and receiver, respectively, and the conversion probability is [11,12,14]

$$p = \frac{\omega}{k_a} \left[ \frac{1}{2} g_{a\gamma\gamma} B_0 L \frac{\sin qL/2}{qL/2} \right]^2. \tag{4}$$

Here $q = k_\gamma - k_a$ indicates the phase mismatch between the photon and axion fields.

The efficiency of generating axions and regenerating photons can be greatly increased by adding electromagnetic resonators, as shown in Figure 1(b) [4,15]. In this figure a laser is used to generate the incident photons, and mirrors are used to cause the light to pass through the magnetic field multiple times, increasing the conversion probability by the factor $2F_T/\pi$, where $F_T$ is the finesse of the resonator in the axion generator (transmitter). This factor can be interpreted as the effective number of photon passes in the resonator. Axions are also emitted in the backward direction owing to the counter-propagating light in the resonator, resulting in half the particles traveling in an unwanted direction. Consequently, the probability of conversion in a given direction is increased by the factor $F_T/\pi$. As also shown in Figure 1(b), a resonator on the photon regenerator (receiver) likewise increases the axion-photon conversion probability [15]. Since regenerated photons will be emitted in both directions, detectors are placed on both ends of the receiving optical resonator, enhancing the photon regeneration probability by the factor $2F_R/\pi$, where $F_R$ is the finesse of the receiving resonator. (If the power from a single end is collected, the factor would be $F_R/\pi$, as with the transmitter.) Finally, to turn this into a communication system, we add appropriate electromagnetic wave modulators and detectors as shown in Figure 1(b).

The conversion loss equation (3) is valid when the transmitter and receiver are sufficiently close together that beam diffraction can be neglected, and when the transmitter and receiver have equal cross-sectional areas. For signaling over long distances, neither assumption will be valid in general. To treat the long-distance case, we first calculate the radiated axion field, then calculate the regenerated photons resulting when the radiated field reaches the receiver.

The general solution to Eq. (2) is given by [12]

$$a(\vec{r}) = i\omega g_{a\gamma\gamma} \int_V d^3r' \frac{e^{ik_a|\vec{r}-\vec{r}'|}}{4\pi|\vec{r}-\vec{r}'|} \vec{A}(\vec{r}')\cdot\vec{B}_0(\vec{r}') . \tag{5}$$

If the observation point $\vec{r}$ is very far away from all points in the source volume $V$, then we obtain the far-field approximation for the axion field

$$a(\vec{r}) \approx i\omega g_{a\gamma\gamma} \frac{e^{ik_a r}}{4\pi r} \int_V d^3r' e^{-i\vec{k}_a\cdot\vec{r}'} \vec{A}(\vec{r}')\cdot\vec{B}_0(\vec{r}') . \tag{6}$$

Consider the case where the source $\vec{A}(\vec{r}')\cdot\vec{B}_0(\vec{r}')$ is only nonzero inside a cylinder of radius $R$ and length $L$ as shown in Figure 2. Further, we assume that within this cylinder $\vec{B}_0 = \hat{x}B_{0T}$ and $\vec{A} = \hat{x}\sqrt{F_T/\pi}A_0\exp(ik_\gamma z)$, where $A_0$ is the amplitude of the incident electromagnetic wave, and the cylinder is contained in a resonant cavity with finesse $F_T$. Using Eq. (6), the far-field potential is found to be

$$a(\vec{r}) \approx i\omega g_{a\gamma\gamma} \frac{e^{ik_a r}}{2\pi r} \sqrt{\frac{F_T}{\pi}} A_0 B_{0T} L_T s_T \frac{\sin\left[(k_\gamma - k_{az})L_T/2\right]}{(k_\gamma - k_{az})L_T/2} \left[\frac{J_1(k_{a\rho}R_T)}{k_{a\rho}R_T}\right], \tag{7}$$

where $k_{a\rho} = k_a \sin\theta$, $k_{az} = k_a \cos\theta$, and $s_T = \pi R_T^2$ is the cross-sectional area of the source region at the transmitter. This expression has its maximum when $\theta = 0$, for which $k_{az} = k_a$, $k_{a\rho} = 0$. The factor containing the Bessel function in (7) has the limit $\lim_{k_{a\rho}\to 0}\left[J_1(k_{a\rho}R_T)/k_{a\rho}R_T\right] = 1/2$. Consequently, the axion field on axis is

$$a(\vec{r}) \approx i\omega g_{a\gamma\gamma} \frac{e^{ik_a r}}{4\pi r}\sqrt{\frac{F_T}{\pi}} A_0 B_{0T} L_T s_T \frac{\sin[qL_T/2]}{qL_T/2}. \tag{8}$$

The time averaged transmitted power density is

$$\langle S_a(\vec{r})\rangle = \frac{1}{2}\omega k_a |a(\vec{r})|^2 = s_T \frac{k_a^2}{(2\pi r)^2}\frac{F_T}{\pi} p_T P_{in}, \tag{9}$$

where $P_{in} = \langle S_\gamma(0)\rangle s_T$, and $\langle S_\gamma(0)\rangle = \frac{1}{2}\omega k_\gamma |A(0)|^2$. If this axion power density is incident upon a photon regenerator at distance $r$ that is perfectly aligned with the transmitter, then the received power is

$$P_0 = (2F_R/\pi)p_R \langle S_a(\vec{r})\rangle s_R. \tag{10}$$

Substituting Eq. (9) for the power flux $\langle S_a(\vec{r})\rangle$ gives

$$P_0 = \frac{P_{in}}{4\pi r^2}\frac{2F_R}{\pi}p_R s_R \frac{k_a^2}{\pi}\frac{F_T}{\pi}p_T s_T. \tag{11}$$

This expression can be understood in terms of antenna theory for electromagnetic waves. In this context, we refer to the apparatus consisting of the resonator and the structure creating the magnetic field as an axion antenna. In analogy with conventional antenna theory, we define the directivity as

$$D = \frac{\langle S_a(\vec{r})\rangle}{P_{rad}/(4\pi r^2)}, \tag{12}$$

where $P_{rad}$ is the total power radiated by the transmitting antenna. To obtain the total power radiated, we could integrate the power flux (9) over a sphere enclosing the antenna. However, it is easier to do the calculation in the near field using the axion field at the aperture of the transmitting antenna. Using $\langle S_a(L_T)\rangle = (F_T/\pi) p_T \langle S_\gamma(0)\rangle$, we have

$$P_{rad} = 2\langle S_a(L_T)\rangle s_T = (2F_T/\pi) p_T P_{in}. \tag{13}$$

Substituting (13) for the total radiated power and (9) for the power flux, the directivity simplifies to

$$D_T = (2\pi/\lambda_a^2) s_T = (4\pi/\lambda_a^2)(s_T/2). \tag{14}$$

The relation between the directivity and physical area (14) is ½ that found in conventional antenna theory, or equivalently, the area appears to be half the physical area. This results from the bi-directional radiation properties of the resonator. Defining an efficiency as $\eta = P_{rad}/P_{in}$, we also define the antenna gain as

$$G_T = \eta_T D_T = (2F_T/\pi) p_T (2\pi/\lambda_a^2) s_T, \tag{15}$$

where $\eta_T = (2F_T/\pi) p_T$.

Next suppose that at some distant location this transmitted field is incident upon a receive antenna with length $L_R$, radius $R_R$, and finesse $F_R$. From (10) and assuming the photons emitted from both ends of the receive antenna are collected, the total power collected will be $P_0 = s_{e,R}\langle S_a(\vec{r})\rangle$, where we have defined the effective area of the receiving antenna as

$$s_{e,R} = s_R(n_c F_R/\pi) p_R, \tag{16}$$

$s_R = \pi R_R^2$ is the physical cross-sectional area, and $n_c$ is the number of ends from which photons are collected (i.e., $n_c = 1, 2$).

As with electromagnetic antennas, the ratio of effective area to gain is found to be independent of the details of the antenna, other than whether or not photons are collected from both ends of the antenna when used to receive:

$$s_{e,R} / G_R = s_{e,T} / G_T = n_c \lambda_a^2 /(4\pi). \tag{17}$$

If photons were collected only from one end of the receive antenna ($n_c = 1$), then Eq. (17) would be identical to conventional antenna theory.

With these definitions, the expression for the received power (11) can be interpreted as a Friis-like equation:

$$P_0 / P_{in} = n_c G_T G_R \left(\lambda_a / 4\pi r\right)^2 = \left(s_{e,T} /(n_{c,T}\lambda_a^2)\right)\left(s_{e,R} / r^2\right). \tag{18}$$

Here $n_{c,T}$ is the value used to compute $s_{e,T}$ according to Eq. (16). The ratio $s_{e,T} / n_{c,T}$ is independent of the choice of $n_{c,T}$, as it should be since the number of detectors that might be used on receive is independent of the transmit properties of the antenna.

It is also useful to note that if the magnetic field is uniform (i.e., wigglers, or quasi-phase matching, are not used [12]), then there is an optimum length for the conversion region of an antenna. This occurs when $\sin(qL/2) = 1$, or $qL = \pi$. The optimum length is found to be

$$L_{opt} = \lambda_\gamma (\omega / m_a)^2. \tag{19}$$

In obtaining this expression, we have used $q \approx m_a^2/(2\omega)$, which is valid for $m_a \ll \omega$.

The diffraction-limited power pattern of the radiated axion field is determined by the aperture size in wavelengths through the Bessel function term in (7):

$$P_d(\theta) = 4\left[J_1(k_a R \sin\theta)/(k_a R \sin\theta)\right]^2. \tag{20}$$

The diffraction beam width between first nulls is determined by the first zero of the Airy disc, and for small angles is given by the well-known expression

$$\theta_{FWFN}^d \approx 1.22 \lambda_\gamma / R_T. \tag{21}$$

Similarly, the diffraction beam width at half maximum is determined by the roots of $P_d(\theta) = 1/2$, or

$$\theta_{FWHM}^d = (1.616/\pi)\lambda_a / R_T \approx 0.514 \lambda_\gamma / R_T. \tag{22}$$

In contrast, the conversion-limited power pattern is given by

$$P_c(\theta) = \left[\frac{\pi}{2}\frac{\sin\left[(k_\gamma - k_a \cos\theta)L/2\right]}{(k_\gamma - k_a \cos\theta)L/2}\right]^2, \tag{23}$$

and depends on both the length in wavelengths and the velocity mismatch. For the optimum length $L$ given by (19), the conversion beam widths are approximately given by

$$\theta_{FWFN}^c \approx 2(m_a/\omega), \qquad \theta_{FWHM}^c \approx 1.06(m_a/\omega). \tag{24}$$

For quantum-limited detection, the channel capacity is [16]

$$C = \Delta\nu \log_2(1 + \eta_d N_R / \Delta\nu), \tag{25}$$

were $\eta_d$ is the quantum efficiency of the optical detectors, $\Delta\nu$ is the bandwidth of the resonator $\Delta\nu = c/(2LF)$ [17], and $c$ is the velocity of light. Equation (25) can be combined with (11) or (18) to find the axion parameters that would permit a particular channel capacity at a given distance for a particular experimental apparatus.

As an example, consider transmitters and receivers with $\lambda_\gamma = 1064$ nm (1.17 eV), $P_{in} = 10$ W, $s = 0.01$ m$^2$, $B_0 = 10$ T, $L = 3$ m, $F = 3.1\times 10^5$, $\eta_d = 0.5$, and a minimum information capacity of 1 bps. Figure 3 shows the inverse coupling strength curves $M(m_a) = 1/g_{a\gamma\gamma}$ for distances of 1000 km and the diameter of the earth. Also shown is the curve for communication between the earth and the far side of the moon using the "4+4" experimental apparatus proposed in [15]. The shaded regions are excluded by the results from the BFRT collaboration [13] and Robilliard et al. [8]. The dots represent extensions of the curves if quasi-phase matching (QPM) is used by periodically reversing the magnetic field [12]. The minimum period of the reversal is taken to be twice the beam diameter for the 3 m system example, and 28.6 m for the 4+4 system. From the figure, communication over distances in excess of 1000 km should be possible for $m_a < 3.5$ meV and $M < 2\times 10^7$ GeV ($g_{a\gamma\gamma} > 5\times 10^{-8}$ GeV$^{-1}$). Of particular interest, we note that the range $2\times 10^6$ GeV $< M < 2\times 10^7$ GeV has not yet been excluded by photon regeneration experiments.

For an area of 0.01 m$^2$, the radius is $R = \sqrt{0.01/\pi} = 0.0564$ m. The half-power diffraction beam width for the example antennas is therefore $\theta^d_{FWHM} = 9.7 \times 10^{-6}$ rad $\approx 5.56 \times 10^{-4}$ deg $\approx 2$ arc-sec, while the half-power conversion beam width for $m_a \approx 0.7$ meV (optimum for $L=3$ m) is $\theta^c_{FWHM} = 6.3 \times 10^{-4}$ rad $\approx 3.63 \times 10^{-2}$ deg $\approx 131$ arc-sec. Consequently the total beam width is determined by the diffraction beam width in this example. While pointing with an accuracy of 2 arc-sec would be challenging, it is somewhat less stringent than that required for optical deep space communications [18]. The size of such axion transmitters and receivers would be roughly that of a medium size professional telescope. It is interesting to note that the size of the beam at a distance of the earth's diameter is about 124 m. Consequently coordinates obtained with differential GPS at both sites should enable the computation of pointing directions to sufficient accuracy.

Note that the receive resonator must be tuned to the same frequency as the transmit resonator to within a fraction of the resonator line width which is 161 Hz in this example. This will present a significant challenge, especially since the resonators are remote from one another. A possible approach would be to use atomic clocks to stabilize both the transmit frequency and a local reference frequency at the receiver. The receive resonator would then be locked to the reference to get as close as possible to the correct frequency, then slowly tuned until the signal is located. A feedback loop could then be closed to lock the receive resonator to the signal.

In summary, for $m_a < 3.5$ meV and $M < 2 \times 10^7$ GeV ($g_{a\gamma\gamma} > 5 \times 10^{-8}$ GeV$^{-1}$), it may be possible to realize a new type of wireless signaling that cannot be blocked or shielded. An example calculation shows that communication between points located diametrically opposite on the earth should be possible. This could enable world-wide communication without the use of satellites or the ionosphere. However, with present knowledge, the signaling will be limited to low data rates, perhaps on the order of a few bits per second for terrestrial links. This estimate assumes 3 m long generation/regeneration regions to allow fully steerable instruments. Though not easily steerable, the apparatus in the "4+4" experiment proposed by Sikivie et al., [15] may enable communication to the far side of the moon for $m_a < 0.3$ meV and $M < 6 \times 10^6$ GeV.

I would like to acknowledge helpful discussions with Jim Lesh, Rich Holman, Jeff Peterson, Pierre Sikivie, and David Tanner during the development of these ideas.

Electronic address: stancil@cmu.edu

Figure Captions

**Figure 1. (a) Basic system for the generation and detection of axions. (b) An axion communication system, with the axion generator and photon regenerator located remote from one another.**

**Figure 2. A source with uniform amplitude over a cylindrical region.**

**Figure 3. (Color Online) The ranges of inverse coupling parameter $M$ and pseudoscalar mass $m_a$ that would permit communication at information bandwidths of at least 1 bps using the example system, and the 4+4 system proposed in [15]. The shaded regions are ruled out by [8] and [13].**

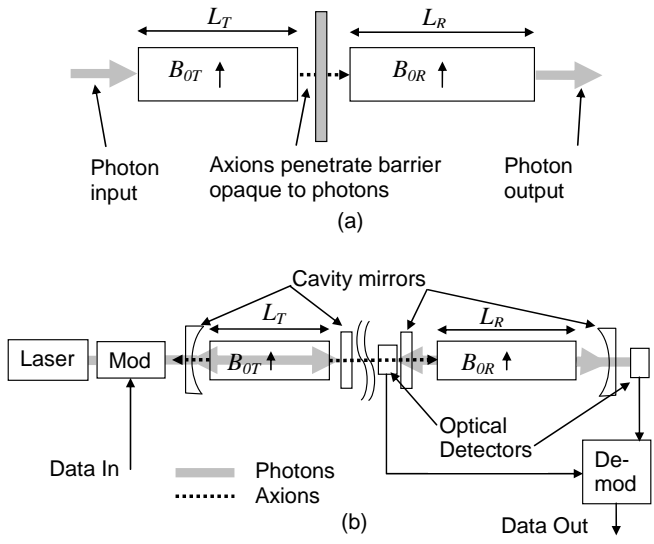

Figure 1, Stancil, Phys. Rev. D.

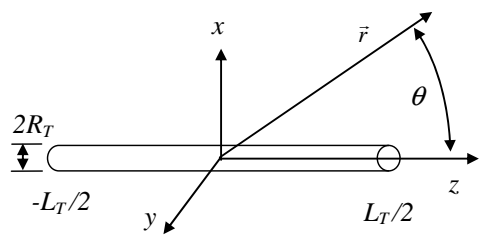

Figure 2, Stancil, Phys. Rev. D.

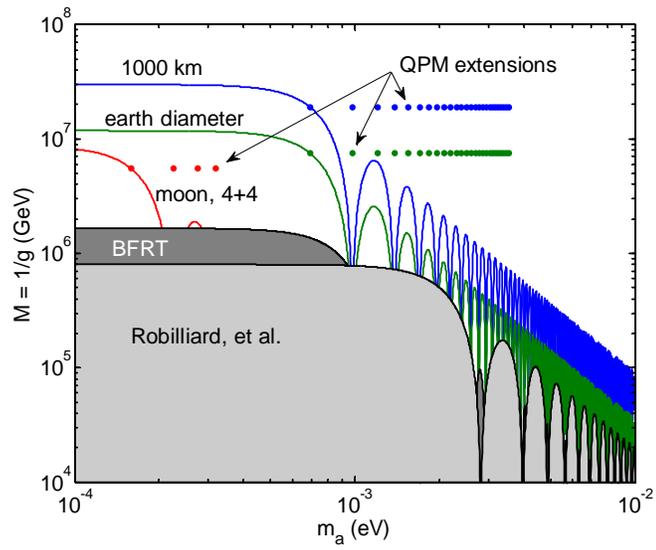

Figure 3, Stancil, Phys. Rev. D.